\documentclass[twocolumn]{jpsj2}
\def\runauthor{Satoshi \textsc{Fujimoto} 
and Norio \textsc{Kawakami}}

\title{%
Kondo Problem and Related One-Dimensional Quantum Systems:\\
Bethe Ansatz Solution and Boundary Conformal Field Theory 
}

\author{%
Satoshi Fujimoto\thanks{E-mail address: fuji@scphys.kyoto-u.ac.jp}
and Norio Kawakami $^1$}

\inst{%
Department of Physics, Kyoto University, Kyoto 606-8502, Japan \\
$^1$Department of Applied Physics, Osaka University, Suita, Osaka 565-0821, Japan
}

\recdate{\today}

\abst{%
We review some exact results on Kondo impurity systems derived from
Bethe-ansatz solutions and boundary conformal field theory
with particular emphasis on universal aspects of the phenomenon.
The finite-size spectra characterizing the low-energy fixed point 
are computed from the Bethe-ansatz solutions
of various models related to the Kondo problem. 
Using the finite-size scaling argument, 
we investigate their exact critical properties.
We also discuss that a universal relation between the Kondo effect 
and the impurity effect in  one-dimensional quantum
systems usefully expedites 
our understanding of these different phenomena. 
}

\kword{ Kondo effect, Bethe ansatz, boundary conformal field theory
}

\begin{document}
\maketitle

\section{Introduction}

A magnetic impurity introduced into metals drastically changes
the local electronic structure of the host systems
through quenching processes of the local magnetic moment;
the phenomenon, known as the Kondo effect, is one of the most
well-understood many-body problems 
in condensed matter physics.\cite{kon,yosi}
This old issue is still a topic of great interest
in connection with modern subjects in the field such as
the exploration of non-Fermi liquid states in heavy fermion compounds and 
the application to nanostructure devices like quantum dots.

In the development of theoretical investigations,
the Bethe ansatz (BA) exact solutions have played an important role,
yielding numerous non-perturbative results on thermodynamic 
quantities.\cite{bethe,kawa}
The BA methods successfully elucidated strong-coupling
aspects of the phenomenon.  
A decade ago, another important breakthrough was achieved by
Affleck and Ludwig in the application of 
 boundary conformal field theory (CFT) to the Kondo problem.\cite{aff1}
CFT is a powerful tool to analyze exact low-energy critical
properties including multi-point correlation functions,
which are not obtained by the BA methods.
In particular, the CFT analysis of the multi-channel Kondo problem 
revealed anomalous critical properties of the non-Fermi 
liquid state.\cite{aff1} 
These two theoretical methods have complementarily 
advanced our understanding on
the Kondo effect and the related impurity problems in one-dimensional
(1D) quantum systems. In this paper, we give a brief overview 
of some recent developments in these directions.
 
The CFT analysis is based upon an effective low-energy theory,
which may be derived from the microscopic lattice Hamiltonians.
In the Kondo problem, however, the derivation of the low-energy effective 
field theory is not straightforward because of its
strong-coupling character. Affleck and Ludwig postulated
that the universality class at the low-energy fixed point belongs to
the SU(2) Kac-Moody CFT.
The validity of this CFT analysis was established by
the consistency with the results obtained by other methods.
Using the BA solutions of various models related to
the Kondo problem, we can directly calculate
the exact finite size spectra (FSS), which characterize the universality
class, and compare them with the results of CFT. 
In \S~2, we review some results related to this issue by exploiting
the single-channel Anderson model, the multi-channel Kondo model, 
and the two-channel Anderson model. In \S~3,  we then
discuss various critical properties of these systems 
by applying the CFT analysis
to the exact FSS obtained from the BA solutions. 

Furthermore, these elaborated techniques revealed universal features of
the phenomenon,  clarifying a profound relation
between the Kondo impurity problem and the issue of
impurity effects in 1D quantum systems.
For instance, in \S~4 , we will discuss 
non-magnetic impurity effects in quantum spin chains with
paying special attention to
 its similarity to the underscreening Kondo effect.

%%%%%%%%%%%%%%%%%%%%%%%%%%%%%%%%%%%%%%%%%%%%%%%%%%%%%%%%%%%%%%%%%%%%
\section{Exact Finite Size Spectra  and Boundary CFT}
%%%%%%%%%%%%%%%%%%%%%%%%%%%%%%%%%%%%%%%%%%%%%%%%%%%%%%%%%%%%%%%%%%%

As was elucidated by Affleck and Ludwig, the critical properties
inherent in the Kondo problem are described by boundary CFT.\cite{aff1}
Before discussing  the Kondo problem, 
we briefly summarize the results of boundary CFT.\cite{cardy}

In general, critical behavior near the boundary is 
characterized by the surface critical exponent $x_s$ which controls the
asymptotic behavior of the  correlation function,
$\langle \phi(t) \phi(0) \rangle \sim 1/t^{2x_s}$
($t\rightarrow \infty$). The surface  exponent
$x_s$ is generally different from bulk exponents.
We can regard the surface critical exponent $x_s$
as the conformal dimension of some boundary scaling operator.
Decomposing a scaling field $\phi$ into the holomorphic
part and the anti-holomorphic part, which is defined by the
analytic continuation of the holomorphic part,
$\phi(z, \bar{z})=\phi_L(z)\phi_R(\bar{z})$,
we have $\phi(z, \bar{z})=
\phi_L(x+iy)\phi_L(x-iy)\sim y^{-2\Delta+\Delta_b}\phi_b(x)$.
Here $\Delta$ is the bulk conformal dimension of $\phi_L$, and
$\Delta_b$ is the conformal dimension of a boundary operator
$\phi_b(x)$. Then in the vicinity of the boundary, we have
the correlation function of the scaling operator,
%%%%%%%%%%%%%%%%%%%%%%%%%%%%%%%%%%%%%
\begin{equation}
\langle \phi (z, \bar{z})\phi(z', \bar{z}')\rangle \sim
\langle \phi_b(x)\phi_b(x')\rangle \sim \vert x-x'
\vert^{-2\Delta_b}.
\label{correlate}
\end{equation}
%%%%%%%%%%%%%%%%%%%%%%%%%%%%%%%

According to the finite size scaling analysis in CFT,
the boundary scaling dimension $\Delta_b$ for
a given  boundary operator
in 1D critical systems with open ends enters in the FSS,
%%%%%%%%%%%%%%%%%%%%%%%%%%%%%
\begin{eqnarray}
E=E_0-\frac{\pi v c}{24l}+\frac{\pi v}{l}(\Delta_b+n). \label{fs}
\end{eqnarray}
%%%%%%%%%%%%%%%%%%%%%%%%%%%%%%%%%
Here $l$ is the linear size of the finite system.
Thus, critical exponents are derived from the FSS.
In the case of the usual Kondo effect (the completely screened case), 
the exponents of correlation functions
are characterized by those of the canonical Fermi liquid.
In contrast, for the overscreening case realized 
in the multi-channel Kondo model,
anomalous dimensions appear, signaling the non-Fermi liquid state.
In the following, we show the results of the FSS for some exactly
solvable models related to the Kondo problem.

%%%%%%%%%%%%%%%%%%%%%%%%%%%%%%%
\subsection{Anderson model}
%%%%%%%%%%%%%%%%%%%%%%%%%%%%%%%

We, first, discuss how critical properties of the Kondo
impurity problem are derived from the FSS, by considering
the single-impurity Anderson model as an example.\cite{fky1}
The model describes
free conduction electrons coupled with
correlated $d$-electrons at the impurity site via the resonant
hybridization $V$.
The Hamiltonian is given by,
%%%%%%%%%%%%%%%%%%%%%%%%%%%%%
\begin{eqnarray}
H&=&\sum_{k\sigma}\varepsilon_kc^{\dagger}_{k\sigma}c_{k\sigma}
+V\sum_{k\sigma}(c^{\dagger}_{k\sigma}d_{\sigma}+
d^{\dagger}_{\sigma}c_{k\sigma}) \nonumber \\
&&+\varepsilon_d\sum_{\sigma}d^{\dagger}_{\sigma}d_{\sigma}
+Ud^{\dagger}_{\uparrow}d_{\uparrow}d^{\dagger}_{\downarrow}d_{\downarrow}
\end{eqnarray}
%%%%%%%%%%%%%%%%%%%%%%%%%%%
with standard notations. 
The Hamiltonian is exactly solvable in terms of the BA method
under the condition that the density of states for conduction
electrons is constant around the Fermi energy.\cite{bethe,kawa}
We can obtain the FSS by using standard techniques
in BA,
%%%%%%%%%%%%%%%%%%%%%%%%%%%%%%%%%%%%%%
\begin{eqnarray}
E&=&E_0+\frac{1}{L}E_1+\frac{1}{L^2}E_2+O(1/L^3), \label{eqn:fss1}\\
\frac{1}{L}E_1&=&\frac{2\pi v}{L}[\frac{1}{4}(\Delta N_h
-2\frac{\delta_F}{\pi})^2+n^{+}_c] \nonumber \\
&&+\frac{2\pi v}{L}[(\Delta S_h)^2+n^{+}_s], \label{eqn:fss2}\\
\frac{1}{L^2}E_2&=&\frac{2\pi v}{L^2}\frac{\chi^{imp}_c}{\chi^h_c}
[\frac{(\Delta N_h)^2}{4}+n^{+}_c]  \nonumber \\
&&+\frac{2\pi v}{L^2}\frac{\chi^{imp}_s}{\chi^h_s}
[(\Delta S_h)^2+n^{+}_s], \label{eqn:fss3}
\end{eqnarray}
%%%%%%%%%%%%%%%%%%%%%%%%%%%%%%%%%%
where $\chi^{imp}_{c (s)}$ and $\chi^h_{c (s)}$ are the charge (spin)
susceptibility of impurity and host electrons, respectively.
$\Delta N_h$ ($\Delta S_h$) is the deviation of
the total number (magnetization) of host electrons
from the ground-state values. 
$n^{+}_{c(s)}$ is a non-negative integer, which features the 
conformal tower structure.
The expression for the $1/L$ correction coincides with the 
fusion hypothesis, supporting the boundary CFT analysis.
The $1/L^2$-term (\ref{eqn:fss3}) characterizes non-universal
correlation effects depending on the strength of $U$.
Namely, the local spin and charge susceptibilities can be 
obtained from the $1/L^2$ corrections of the FSS.

To read off boundary scaling dimensions 
from eqs.(\ref{eqn:fss1}) and (\ref{fs}),
we replace $L$ with $2l$, since $L$ is defined as
periodic length of the system.
Note that the finite-size spectrum for bulk
electrons in eq.(\ref{eqn:fss2}) involves a
non-universal phase shift $\delta_F$. 
The phase shift $\delta_F$ is regarded as
the chemical-potential change due to the
impurity.
This means that the effect of the phase
shift amounts to merely imposing twisted boundary conditions
on conduction electrons \cite{aff1}.
Therefore, when we derive the dimension
of a scaling operator associated with conduction electrons, we
should discard  the $\delta_F$ dependence
in eq.(\ref{eqn:fss2})
by redefining $\Delta N_h-2\delta_F/\pi\rightarrow\Delta N_h$.
Hence the scaling dimension $x$ of the conduction electron field
is obtained by taking the quantum numbers
%%%%%%%%%%%%%%%%%%%%%%%%%%%
\begin{equation}
\Delta N_h=1, \hskip 5mm \Delta S_h= 1/2,
\label{eqn:quanta}
\end{equation}
%%%%%%%%%%%%%%%%%%%%%%%%%%%
resulting in $x=1/2$.
Thus we can see that the single-electron Green function
$\langle c_{\sigma}(t)c^{\dagger}_{\sigma}(0)\rangle\sim 1/t^{\eta}$
has the canonical exponent $\eta=2x=1$, in accordance
 with the fact that the
system is described by the strong-coupling fixed point of
the local Fermi liquid \cite{fermi,yamada,aff1}.

On the other hand, we can extract another interesting
information from the FSS
(\ref{eqn:fss2}),  i.e. the critical behavior
related to the orthogonality catastrophe.
To see this explicitly, let us consider the
time-dependent Anderson model in which
the hybridization $V=0$
for $t<t_0$, and then $V$ is switched on at $t=0$.
This system shows the orthogonality catastrophe related
to the Fermi edge singularity.
We note that 
the model is defined on a semi-infinite plane with the boundary at $x=0$.
Using the conformal transformation $t+ix=\exp[(\pi/l)(t'+ix')]$,
we map the semi-infinite plane to a strip with width $l$,
in which the free boundary condition is imposed at $x'=l$, and,
contrastively, the non-trivial Kondo boundary condition is imposed at
$x'=0$. 
We recall that the FSS (\ref{eqn:fss2}) is computed 
under these boundary conditions.
Thus, critical exponents which govern the long-time behavior of 
the correlation functions  characterizing the orthogonality 
catastrophe can be obtained from  eq.(\ref{eqn:fss2}). 
In this case, the phase shift in
(\ref{eqn:fss2}) becomes  a key quantity
which controls the critical exponent. 
Actually, critical
exponents related to the X-ray problem can be read off from
the above FSS with keeping the dependence on the
phase shift intact \cite{fky1,afflud}.
This is because a sudden potential change occurs
in X-ray photoemission (or absorption) experiments.
For example, the critical exponent $\eta$ of
the single-particle Green function
%%%%%%%%%%%%%%%%%%%
$\langle c_{\sigma}(t)c_{\sigma}^{\dagger}(0)\rangle
\sim 1/t^{\eta}$ $(0<t_0 \ll t)$
%%%%%%%%%%%%%%%%%%%
for the  model with a sudden potential change
is obtained as \cite{afflud}
%%%%%%%%%%%%%%%%%%%%%%%%%%%%%%%%%%
\begin{equation}
\eta=1-\frac{2\delta_F}{\pi}+2\biggl(\frac{\delta_F}{\pi}\biggr)^2.
\end{equation}
%%%%%%%%%%%%%%%%%%%%%%%%%%%%%%%
This is just the exponent which governs the
long-time behavior of the overlap integral  between
the initial and final states in
the X-ray absorption problem.
It is thus seen that the FSS (\ref{eqn:fss2})
contains the information about orthogonality catastrophe
in addition to the local Fermi-liquid properties.

%%%%%%%%%%%%%%%%%%%%%%%%%%%%%%%%%%%%%%%%
\subsection{Multi-channel Kondo model}
%%%%%%%%%%%%%%%%%%%%%%%%%%%%%%%%%%%%%%%

The argument in the previous section is also applicable to
the multi-channel Kondo model.\cite{nb,tw,andrei}
In this model, conduction electrons with  $n$-channels ($l=1, 2, \cdots, n$)
couple with the impurity spin $S$ antiferromagnetically.
In the case of $n>2S$, the overscreening Kondo effect occurs,
yielding the non-Fermi liquid ground state.
The Hamiltonian reads
%%%%%%%%%%%%%%%%%%%%%%%%%%%%%%%%
\begin{eqnarray}
H=\sum_{km\sigma}\varepsilon_kc^{\dagger}_{km\sigma}c_{km\sigma}
+J\sum_{kk'm\sigma\sigma'}c^{\dagger}_{km\sigma}(\sigma_{\sigma\sigma'}\cdot
\mbox{\boldmath $S$}) c_{k'm\sigma'},
\end{eqnarray}
%%%%%%%%%%%%%%%%%%%%%%%%%%%%%%%%%%%%%
where $c_{km\sigma}$ is the annihilation operator for electrons with
orbital index $m$ and spin $\sigma$. 
It is noted that although the multi-channel Kondo model is
BA-solvable,\cite{tw,andrei} we cannot apply 
standard techniques to the calculation of the FSS in this model,
particularly, in the overscreening case $n>2S$.
This is due to the fact that
the BA solution to the multi-channel model 
takes the form of the so-called string solution 
even for the ground state.\cite{tw,andrei} 
It has been known that the string solution is valid only in the
thermodynamic limit. Thus,
a naive application of finite-size techniques
\cite{finite} fails, giving only the Gaussian part of the
spectrum  for the spin sector.
Since the spin sector of the multi-channel Kondo model
is described by the level-$n$ SU(2) Kac Moody theory
with the central charge $c_{WZW}=3n/(2+n)$ \cite{aff1},
the $Z_n$ parafermion sector with $c=c_{WZW}-c_{\rm Gaussian}=2(n-1)/(n+2)$
is not accessible by standard analytical techniques. 
%It is thus necessary to exploit alternative methods other than the
%coordinate BA to obtain the correct spectrum corresponding to
%the missing Z$_n$ parafermions.
To overcome the difficulty, we propose an alternative method
to calculate the FSS analytically with paying a special attention
to the nontrivial Z$_n$ parafermion part.\cite{fk2}
To this end, we extract the Z$_n$ sector of the BA equations 
for the multi-channel model in the overscreening case $n>2S$,
%%%%%%%%%%%%%%%%%%%%%%%%%%%%%%%
\begin{eqnarray}
\tilde{f}_m+A^{(n)}_{m,k}*f_k&=&A^{(n)}_{m,n-1}*s*\tilde{f}_n \nonumber \\ 
&&+\frac{1}{L}A^{(n)}_{m,2s}*s(\lambda+1/J), \label{zn}
\end{eqnarray}
%%%%%%%%%%%%%%%%%%%%%%%%%%%%%%
where $*$ means the convolution.
Here $f_m$ and $\tilde{f}_m$ are the distribution functions for
rapidities of particle and hole excitations, respectively.
The kernel is defined as $A^{(n)}_{mk}=(C^{(n)})^{-1}_{mk}$ with
%%%%%%%%%%%%%%%%%%%%%%%%%%%%%%%%%%%%%%%%%%%%%%%%%%%%%%%%%%%%
\begin{eqnarray}
C_{mk}^{(n)}=\delta_{m,k}-s(\lambda)(\delta_{m,k+1}+\delta_{m,k-1}),
\end{eqnarray}
%%%%%%%%%%%%%%%%%%%%%%%%%%%%%%%%%%%%%%%%%%%
and $s(\lambda)=1/(2\cosh(\pi\lambda))$.
Note that the second term of the right-hand side of eq.(\ref{zn})
reflects the existence of the impurity, and appears only for $n>2S$.
The bulk part of eq.(\ref{zn}) gives the exact spectrum of
the Z$_{n}$ parafermion theory.
We recall the fact that this BA equation is also derived from
the S-matrix for ``physical particles'' of the Z$_{n}$ model.\cite{fend}
%the bulk S-matrix for ``physical particles'' \cite{fend}
%in the spin sector for the overscreening case
%is decomposed into two parts\cite{fend};
%the S-matrix for the Gaussian model  and that for the Z$_{n}$ model.
Since the Z$_{n}$ model can be described by
the restricted solid-on-solid (RSOS)
model in the regime I/II\cite{abf},
the corresponding S-matrix is given by the face weight
in the RSOS model \cite{res}.
A remarkable point for the overscreening model is that
the interaction between ``physical particles''
and the impurity is described by the S-matrix
of multi-kinks\cite{fend},
which is given by the fusion of the face
weights of the RSOS model with the
fusion level $p=n-2S$.
Therefore, the spectrum of eq.(\ref{zn}) is
essentially determined by the RSOS model coupled with the impurity, of
which the FSS can be obtained analytically
by using the functional equation method\cite{kp}.
By some technical reasons, we restrict our analysis to the case
of $p\equiv n-2S=1$ here.  After some manipulations,
we end up with  the FSS of the RSOS model coupled
with the impurity\cite{fk2},
%%%%%%%%%%%%%%%%%%%%%%%%%%%%%%%%%%%%%%
\begin{eqnarray}
E_{RSOS}=\frac{2\pi v}{L}\biggl(\frac{j(j+1)}{n+2}
-\frac{(m+p)^2}{4n}\biggr)+\mbox{const.} \label{fssrsos}
\end{eqnarray}
%%%%%%%%%%%%%%%%%%%%%%%%%%%%%%%%%%%%%%
where $ m=2j\quad(\mbox{mod} \,\, 2)$, and
$j=0, 1/2, 1, ..., n/2$.
It is seen that the spectrum fits in with
Z$_n$ parafermion theory, and only the selection rule for
quantum numbers is changed
by the impurity effect, $m \rightarrow (m+p)$.
Adding eq.(\ref{fssrsos}) to the Gaussian spectrum, 
we arrive at the total FSS,
%%%%%%%%%%%%%%%%%%%%%%%%%%%%%%%%%%%%%%
\begin{eqnarray}
& E&=\frac{2\pi v}{L}\biggl(\frac{(Q-n)^2}{4n}
+\frac{\tilde{j}(\tilde{j}+1)}{n+2}
+(\mbox{orbital part}) \nonumber \\
&&+n_Q+n_s+n_f\biggr)
\end{eqnarray}
%%%%%%%%%%%%%%%%%%%%%%%%%%%%%%%%%%%%%%%
where $\tilde{j}=|j-p/2|$ is the new quantum number.
Here $Q$ and $j$ are the charge and spin  quantum
numbers for free electrons without the Kondo impurity.
$n_Q$, $n_s$, and $n_f$ are non-negative integers
characterizing the conformal tower.
We can say that the effect
due to the Kondo impurity is merely to modify the selection rule for
quantum numbers of spin excitations by
$j \rightarrow \tilde j$,
which indeed results in non-Fermi liquid state.
This is the essence of the Kac-Moody fusion hypothesis proposed
by Affleck and Ludwig\cite{aff1}.
The above result may be a microscopic description of the fusion
hypothesis for the multi-channel Kondo model.

%%%%%%%%%%%%%%%%%%%%%%%%%%%%%%%%%%%%%%%%%%%%%%%%%%%%
\subsection{Two-channel Anderson  model}
%%%%%%%%%%%%%%%%%%%%%%%%%%%%%%%%%%%%%%%%%%%%%%%%%%%

Recently, the BA exact solution of the
 two-channel Anderson model
was obtained by Bolech and Andrei.\cite{ba}
This model was proposed to describe a mixed valence regime
of the Uranium compounds such as UBe$_{13}$ that have 
both magnetic and quadrupolar degrees of freedom.\cite{sch}
The model consists of the impurity $f$-electrons,
which are assumed to be
either in the $5f^3$ configuration or in the $5f^2$ configuration.
The former state is the $\Gamma_6$ magnetic spin doublet, and
the latter state is the $\Gamma_3$ quadrupolar doublet. 
Conduction electrons, which carry both magnetic and quadrupolar 
quantum numbers, hybridize with the $f$-electrons.
In the strong coupling limit, the model Hamiltonian reads,
%%%%%%%%%%%%%%%%%%%%%%%%%%
\begin{eqnarray}
&H&=\sum_{\alpha\sigma}\int dx c^{\dagger}_{\alpha\sigma}(x)
(-i\frac{\partial}{\partial x})c_{\alpha\sigma}(x)   \nonumber \\
&&+\varepsilon_s\sum_{\sigma}f^{\dagger}_{\sigma}f_{\sigma}
+\varepsilon_q\sum_{\alpha}b^{\dagger}_{\alpha}b_{\alpha} \nonumber \\
&&+V\sum_{\alpha\sigma}\int dx\delta(x)(f^{\dagger}_{\sigma}b_{\bar{\alpha}}
c_{\alpha\sigma}(x)
+c^{\dagger}_{\alpha\sigma}(x)b^{\dagger}_{\bar{\alpha}}f_{\sigma}). 
\label{ma}
\end{eqnarray}
%%%%%%%%%%%%%%%%%%%%%%%%%%%%%%%%%%%%
Here $c_{\alpha\sigma}(x)$ ($c^{\dagger}_{\alpha\sigma}(x)$) is 
the annihilation (creation) operator of a conduction electron
with spin $\sigma=\uparrow\downarrow$ and quadrupolar quantum number
$\alpha=\pm$.
$\bar{\alpha}$ is the conjugate representation of $\alpha$.
$f_{\sigma}$ ($f^{\dagger}_{\sigma}$) is the annihilation (creation) 
operator for the $\Gamma_6$ magnetic spin doublet on the impurity site, 
and $b_{\alpha}$ ($b^{\dagger}_{\alpha}$) is the annihilation (creation)
operator for the $\Gamma_3$ quadrupolar doublet. 
Under the assumption that 
the Hilbert space is restricted to these two configurations,
the constraint $\sum_{\sigma}f^{\dagger}_{\sigma}f_{\sigma}+
\sum_{\alpha}b^{\dagger}_{\alpha}b_{\alpha}=1$ is imposed. 
The system exhibits crossover between magnetic and quadrupolar Kondo
effects depending on the parameter 
$\varepsilon=\varepsilon_s-\varepsilon_q$;
for $\varepsilon-\mu\ll -\Delta=\pi\rho V^2$, the magnetic overscreened
fixed point appears, and for $\varepsilon-\mu\gg \Delta=\pi\rho V^2$
the low-energy properties are characterized by
the quadrupolar overscreening Kondo effect.

Critical properties of this model were extensively discussed by
Johannesson et al. on the basis of boundary CFT\cite{jab}.
Here we briefly summarize their results.
The low-energy universality class of the model (\ref{ma}) is classified 
by examining the FSS, which is obtained exactly by using the method 
explained in the previous section:
%%%%%%%%%%%%%%%%%%%%%%%%%%%%%%
\begin{eqnarray}
E=\frac{2\pi v}{L}\biggl(\frac{1}{8}(Q-4\frac{\delta_F}{\pi})^2
+\frac{j_s(j_s+1)}{4}+\frac{j_f(j_f+1)}{4}\biggr). \label{tafs}
\end{eqnarray}
%%%%%%%%%%%%%%%%%%%%%%%%%%%%%%%
Here $Q$, $j_s$, and $j_f$ are, respectively, 
the charge, spin, and quadrupole quantum numbers.
$\delta_F$ is the phase shift related to the charge 
number of the impurity
site $\sum_{\sigma}f^{\dagger}_{\sigma}f_{\sigma}$.
The quantum numbers obey the selection rule,
$Q=0$(mod 2), $j_s=0$ or $1$, $j_f=1/2$, or, alternatively,
$Q=1$(mod 2), $j_s=1/2$, $j_f=0$ or $1$.
Note that this FSS depends on the parameter $\varepsilon$ only through
t1he phase shift $\delta_F$.
In the magnetic limit $\varepsilon-\mu\ll -\Delta$,
$\delta_F\rightarrow \pi/2$, and in the quadrupolar limit
$\varepsilon-\mu\gg \Delta$, $\delta_F\rightarrow 0$.
This observation implies that
even in the intermediate mixed valence regime,
%although the model shows 
%the crossover between magnetic 
%and quadrupolar Kondo effects with the continuous variation of 
%$\varepsilon$,
the low-energy fixed point is governed by the Kac-Moody fusion rules
for two-channel systems with impurity spin $s=1/2$
proposed by Affleck and Ludwig.\cite{afflud}
The critical properties of the model are derived from the FSS (\ref{tafs}).

%%%%%%%%%%
Before closing this section, a brief comment is in order for the 
dynamically-induced multi-channel Kondo effect.\cite{tatsuya}
Let us imagine a completely screened Kondo impurity with $S>1/2$.
If one of the core electrons, which compose the impurity spin-$S$,
is emitted via photoemission process, the overscreening Kondo effect 
is induced in the excited state, although the complete screening 
is realized in the initial state.  Therefore,
 anomalous power-law 
behavior inherent in the overscreening model may be observed
in the photoemission spectrum at low energies.
This exemplifies the idea of dynamically-induced (or photo-induced)
multi-channel Kondo effect.

%%%%%%%%%%%%%%%%%%%%%%%%%%%%%%%%%%%%%%%%%%%%%%%%%%%%%%%%%%%%%%%%%%%
\section{Critical Exponents of Pseudo-Particles in the Kondo Problem}
%%%%%%%%%%%%%%%%%%%%%%%%%%%%%%%%%%%%%%%%%%%%%%%%%%%%%%%%%%%%%%%%

The exponents related to the orthogonality catastrophe mentioned 
in \S 2.1 also manifest themselves in the dynamical behavior 
of {\it pseudo-particles}, which are introduced to
describe the strong-coupling limit ($U\rightarrow\infty$) of
the Anderson model.
In this limit, the double occupancy of electrons
is forbidden, and thus the Fock space of the impurity electron
can be mapped to that spanned by the slave-boson field 
and the pseudo-fermion field which represent an empty site
and a singly occupied site, respectively. 
The long-time asymptotic behavior of dynamical
correlation functions of these pseudo-particles
can be derived exactly from a boundary CFT analysis.

For example, we consider the $U \rightarrow \infty$ SU($N$) 
Anderson model.\cite{fky2}
The Hamiltonian is given by
%%%%%%%%%%%%%%%%%%%%%%%%%%%%%%%%%%%%%%%%%%
\begin{eqnarray}
&& H=\sum_{m=1}^{N}\int dx c^{\dagger}_{m}(x)(-i\frac{\partial}{\partial x})
c_{m}(x)+\varepsilon_f\sum_{m=1}^{N}f^{\dagger}_mf_m \nonumber \\
&&+V\sum_{m=1}^{N}\int dx\delta(x)(f^{\dagger}_{m}bc_{m}(x)
+c^{\dagger}_{m}(x)b^{\dagger}f_{m}),
\label{sun}
\end{eqnarray}
%%%%%%%%%%%%%%%%%%%%%%%%%%%%%%%%%%%%%%%%%%%%%%
where $b$ and $b^{\dagger}$ are the annihilation and creation 
operators of the slave-boson,
and $f_m$ and $f^{\dagger}_m$ ($m=1,2,...,N$) are
the annihilation and creation operators of
the pseudo-fermion. Since the double occupancy is
forbidden, the constraint $b^{\dagger}b+\sum_m^Nf^{\dagger}_mf_m=1$ 
is imposed. 
The FSS of this model is readily obtained as,
%%%%%%%%%%%%%%%%%%%%%%%%%%%
\begin{equation}
\frac{1}{L}E_1=
\frac{2\pi v}{L}\frac{1}{2}\Delta {\bf M}^{T}
{\cal C}_{f}\Delta {\bf M}
-\frac{\pi v}{L}N
\biggl(\frac{\delta_F}{\pi}\biggr)^2, \label{sunfss}
\end{equation}
%%%%%%%%%%%%%%%%%%%%%%%%%%%
where ${\bf M}^T=(\Delta M^{(1)}_h,...,\Delta M^{(N-1)}_h,
\Delta N_h-N\delta_F/\pi)$. 
$\Delta N_h$ and $\Delta M_h^{(l)}$ are the number of 
charge excitations and spin excitations, respectively.
${\cal C}_{f}$ is the Cartan matrix for the OSp($N$,1) Lie 
superalgebra.\cite{fky2}
It is intriguing that the Cartan matrix in (\ref{sunfss}) reflects 
the hidden higher symmetry of the impurity model (\ref{sun}).
%Here the $N \times N$ matrix
%${\cal C}_{f}$ is given by
%\begin{eqnarray}
%{\cal C}_{f}=
%\left(
%\begin{array}{cccc}
%2     & -1      & \null  &
%            \smash{\lower1.7ex\hbox{\LARGE 0}} \\
%-1     & 2  & \ddots & \null   \\
%\null  & \ddots  &  \ddots     & -1      \\
%        \smash{\hbox{\LARGE 0}}   & \null   &  -1    & 1  \\
%\end{array}
%\right) .
%\end{eqnarray}

Let us now study the long-time behavior of the
Green functions for pseudo-particles;
$\langle f^{\dagger}_{m}(t)f_{m}(0)\rangle\sim t^{-\alpha_f}$, and
$\langle b^{\dagger}(t)b(0)\rangle\sim t^{-\alpha_b}$.
As explained before, when determining canonical exponents
for the local Fermi liquid,
we can neglect the phase shift in eq.(\ref{eqn:fss1}).
In the derivation of critical exponents for pseudo-particles,
however, we must regard the number of impurity electrons
(or phase shift) $n_l=\delta_l/\pi$ as  a quantum number.
Therefore the phase shift plays an essential role to determine the
critical exponents.  For instance, in order to obtain
the Green function of pseudo-fermions,
we take  $\Delta N_h=1$ and
$\Delta M_h^{(l)}=0$ as quantum numbers. We thus obtain
the corresponding critical exponent as,
%%%%%%%%%%%%%%%%%%%%%%%%%%%%%%%%%%%%%%
\begin{equation}
\alpha_f=1-\frac{2\delta_f}{\pi}+N\biggl(\frac{\delta_F}{\pi}\biggr)^2.
\label{pf}
\end{equation}
%%%%%%%%%%%%%%%%%%%%%%%%%%%%%%%%%%%%%%%%%%%%%%%%
In a similar way, the critical exponent $\alpha_b$ for the
slave-boson Green function
can be obtained. Since the
slave-boson expresses a vacancy, it carries neither charge nor spin.
Putting  $\Delta N_h=\Delta M_h^{(l)}=0$, one gets
%%%%%%%%%%%%%%%%%%%%%%%%%%%
\begin{equation}
\alpha_b=N\biggl(\frac{\delta_F}{\pi}\biggr)^2 \label{pb}
\end{equation}
%%%%%%%%%%%%%%%%%%%%%%%%%%%%%%
These exact expressions
for $\alpha_f$ and $\alpha_b$ agree with those obtained
for the $N=1$ and $N=2$ cases\cite{menge,costi},
and take the same form as those in the X-ray problem:
the exponent of pseudo-fermion corresponds to the X-ray
absorption exponent, and that of slave-boson
to the X-ray photoemission exponent.
In the Kondo effect, the Fermi edge singularity shows up
in the intermediate state as pointed out  in the
Anderson-Yuval approach.\cite{ay} The anomalous exponents discussed here
reflect this singularity.

The above argument is also applicable to the two-channel Anderson model
(\ref{ma}), of which the low-energy fixed point is a non-Fermi liquid.
The exponents for the pseudo-fermion $f_{\sigma}$ for this system
is easily read off from eq.(\ref{tafs}) with $Q=1$, $j_s=1/2$, $j_f=0$ and
the phase shift retained in the charge sector,\cite{jab}
%%%%%%%%%%%%%%%%%%%%%%%%%%%%%%%%%%%%%%%%%
\begin{eqnarray}
\alpha_f^{\rm TA}=\frac{5}{8}-\frac{2\delta_F}{\pi}
+4\biggl(\frac{\delta_F}{\pi}\biggr)^2.
\end{eqnarray}
%%%%%%%%%%%%%%%%%%%%%%%%%%%%%%%%%%%%%%%%
Similarly, the exponent for the slave boson $b_{\alpha}$
is obtained as,\cite{jab}
%%%%%%%%%%%%%%%%%%%%%%%%%%%%%%%%%%%%%%%%%%
\begin{eqnarray}
\alpha_b^{\rm TA}=\frac{3}{8}+4\biggl(\frac{\delta_F}{\pi}\biggr)^2.
\end{eqnarray}
%%%%%%%%%%%%%%%%%%%%%%%%%%%%%%%%%%%%%%%%%
These results present a striking contrast to the exponents for
the Fermi liquid state (\ref{pf}) and (\ref{pb}).

%%%%%%%%%%%%%%%%%%%%%%%%%%%%%%%%%%%%%%%%%%%%%%%%%%%%%%%
\section{``Underscreening Kondo effect'' 
due to Non-Magnetic Impurities in 1D Systems}
%in the Vicinity of Open Boundaries in 1D Quantum Systems}
%%%%%%%%%%%%%%%%%%%%%%%%%%%%%%%%%%%%%%%%%%%%%%%%%%%%%%

In this section, we would like to discuss the analogy between
the Kondo problem and the impurity effects in quantum
1D systems.  The Kondo effect is deeply related with 
the effects of spinless impurities in 1D quantum spin systems.
In 1D systems, non-magnetic impurities typically cut the chains, and play 
the role of open boundary conditions.
Remarkably, magnetic correlations in the vicinity of the open ends
show anomalous behavior similar to 
``underscreening Kondo effect''.\cite{de,ess,as,fra,fuji2,fujiegg,furu}

For example, we consider the $s=1/2$ Heisenberg spin chain with open
boundaries, of which the Hamiltonian is,
%%%%%%%%%%%%%%%%%%%%%%%%%%%%%
\begin{equation}
%H=J\sum_{i=1}^N \mbox{\boldmath $S$}_i\cdot\mbox{\boldmath $S$}_{i+1}
H_{XXZ}=J\sum_{i=1}^N [S_{i}^xS_{i+1}^x+S_{i}^yS_{i+1}^y
+\Delta S_{i}^zS_{i+1}^z].
\label{xxz}
\end{equation}
%%%%%%%%%%%%%%%%%%%%%%%%%%%%%%%%%%%%%
The system is exactly solvable in terms of the BA method.
For the isotropic case $\Delta=1$, the BA equation for rapidities
$\lambda_j$ reads,
%%%%%%%%%%%%%%%%%%%%%%%%%
\begin{eqnarray}
\biggl(\frac{\lambda_j+\frac{i}{2}}{\lambda_j-\frac{i}{2}}\biggr)^{2N+1}
\frac{\lambda_j+i}{\lambda_j-i}=-\prod_{l=1}^M
\frac{\lambda_j-\lambda_l+i}{\lambda_j-\lambda_l-i}. \label{baeh}
\end{eqnarray}
%%%%%%%%%%%%%%%%%%%%%%%%%%%%%%%%%%
As was pointed out by de Sa and Tsvelik, this equation has an analogy to
the BA equation for the $s$-$d$ exchange model 
with arbitrary spin $S$,\cite{de}
%%%%%%%%%%%%%%%%%%%%%%%%%%%%%%
\begin{eqnarray}
\biggl(\frac{\lambda_j+\frac{i}{2}}{\lambda_j-\frac{i}{2}}\biggr)^{N}
\frac{\lambda_j+\frac{1}{c}+iS}{\lambda_j+\frac{1}{c}-iS}=-\prod_{l=1}^M
\frac{\lambda_j-\lambda_l+i}{\lambda_j-\lambda_l-i},\label{baek}
\end{eqnarray}
%%%%%%%%%%%%%%%%%%%%%%%%%%%%%%%
where $c=2J/[1-S(S+1)J^2]$. 
The BA equation for the $s=1/2$ Heisenberg chain with open boundaries
is equivalent to that for the Kondo problem with the impurity spin $S=1$
in the strong coupling case $c\rightarrow \infty$.
We can easily check that 
this similarity is also found in the anisotropic case $\Delta\neq 1$.
Thus, it is expected that the underscreening Kondo effect may emerge
in magnetic properties of eq.(\ref{xxz}).
However, this analogy is not complete. 
In the underscreening Kondo effect, the ground state is not a spin singlet,
because of the presence of the partially unquenched impurity spin.
On the other hand, the ground state of 
the Heisenberg spin chain with open boundaries is a spin singlet
in the thermodynamic limit $N\gg J/T$.
It is important that even in this singlet case, 
the boundary effect on eq.(\ref{xxz}) 
gives rise to non-trivial magnetic behavior
as suggested from the above BA equation.
In fact, the boundary spin susceptibility at zero temperature
for the isotropic case obtained from eq.(\ref{baeh})
shows the singular dependence 
on an external magnetic field,\cite{fujiegg,furu}
%%%%%%%%%%%%%%%%%%%%%%%%%%%%%%%
\begin{eqnarray}
\chi_{\rm B}(T=0)=
%\frac{g^2}{4Nh}+\frac{5g^3}{8Nh}+...
%=
\frac{1}{4Nh(\ln(2\pi \alpha/h))^2}
\biggl(1-\frac{\ln\ln(2\pi\alpha/h)}{\ln(2\pi\alpha/h)}
+...\biggr).\label{zero}
\end{eqnarray}
%%%%%%%%%%%%%%%%%%%%%%%%%%%%%%
Here $\alpha=\sqrt{\pi/2}\exp(1/4+\gamma)$ with $\gamma$ the Euler constant.

The divergent behavior of the boundary spin susceptibility is
also seen in its temperature dependence.
Since it is difficult to calculate
the low temperature behavior from the BA equation,
we carry out a perturbative expansion for the effective field theory,
which gives asymptotically exact low-energy properties.
For the boundary problem, it is convenient to express the partition function
by using the transfer matrix $\exp(-LH^c)$,
where $H^c$ is the Hamiltonian defined on the imaginary time axis.
The effective Hamiltonian in the isotropic case $\Delta=1$ is,\cite{luk}
%%%%%%%%%%%%%%%%%%%%%
\begin{eqnarray}
&& H= H_{WZW}+H_m, \label{eff2}\\
&& H_m= -g\int_0^{1/T}\frac{d\tau}{2\pi}\sum_{a=1}^3J^a(\tau)\bar{J}^a(\tau).
\label{int2}
\end{eqnarray}
%%%%%%%%%%%%%%%%%%%%%%%%%%%%%%%%%%%%%%%
Here $H_{WZW}$ is the Hamiltonian of the level $k=1$ SU(2)
Wess-Zumino-Witten model, and $J^a$ ($\bar{J}^a$)
is the left (right) moving current of the level $k=1$ SU(2) 
Kac-Moody algebra.
The running coupling constant $g$ depends on
temperature $T$ and an external magnetic field $h$ through the scaling 
equation,\cite{luk}
%%%%%%%%%%%%%%%%%%%%%%%%%%%%%%%
\begin{eqnarray}
g^{-1}+\frac{1}{2}\ln(g)=-{\rm Re}[\psi(1+\frac{i h}{2\pi T})]
+\ln(\sqrt{\pi/2} e^{1/4}J/T), \label{sca}
\end{eqnarray}
%%%%%%%%%%%%%%%%%%%%%%%%%%%%%%%
with $\psi(x)$ the di-gamma function.
The boundary contributions are obtained from a perturbative expansion 
of the free energy,
%%%%%%%%%%%%%%%%%%%%%%%
\begin{eqnarray}
F=-\frac{T}{N}\ln\langle 0|e^{-LH^c}|B\rangle, \label{free}
\end{eqnarray}
%%%%%%%%%%%%%%%%%%%%%%
with respect to $H_m$.
Here $|B\rangle$ and $|0\rangle$ are the boundary and (bulk) vacuum states, 
respectively.
Up to the lowest order, it is sufficient to consider 
the non-perturbative state of $|B\rangle$, {\it i.e.} 
the conformally invariant boundary state of $H_{WZW}$,
of which the expression in terms of free boson fields
is well-known.\cite{bound}
The perturbation term of (\ref{free}) is non-vanishing
only for the Dirichlet boundary condition, which corresponds to
the situation where the local magnetization
at the boundary is not fixed to a particular value.
Eventually, we obtain the leading term of 
the boundary spin susceptibility,\cite{fujiegg,furu}
\begin{eqnarray}
\chi_{\rm B}=
%\frac{g}{12NT}+\frac{g^2}{12NT}(1-\frac{3\psi''(1)}{2\pi^2})
%+... %\nonumber \\
%= 
\frac{1}{12NT\ln(\alpha/T)}\biggl(1-\frac{\ln\ln(\alpha/T)}{2\ln(\alpha/T)}
+...\biggr), \label{chi2}
\end{eqnarray}
Also, a similar temperature dependence is found in 
the boundary contribution of the specific heat coefficient,\cite{fujiegg,furu} 
\begin{eqnarray}
\frac{C_{\rm B}}{T}=
%\frac{g^2}{2NT}+\frac{5g^3}{4NT}+...= 
\frac{1}{2NT(\ln(\alpha/T))^2}\biggl(1-\frac{\ln\ln(\alpha/T)}{\ln(\alpha/T)}
+...\biggr),
\label{spe2}
\end{eqnarray}
We would like to stress that the divergent behavior at low temperatures
appears even in the thermodynamic limit $N\gg J/T$; {\it i.e.} 
irrespective of whether $N$ is even or odd.
These results are also confirmed by the numerical transfer matrix 
renormalization group method.\cite{fujiegg}

The above singular temperature dependence implies that
in the vicinity of open boundaries, spin excitations are very sensitive to
thermal perturbations.
This sensitivity partially stems from a boundary entropy perturbed
by bulk interactions.
The presence of the ground state degeneracy caused by open boundaries
is a universal feature of boundary critical phenomena in 1+1 dimensional 
systems.\cite{egg}
For the $s=1/2$ Heisenberg spin chain with free open boundaries.
the boundary entropy is equal to $\ln\langle 0 |B\rangle=\ln(K/2)^{1/4}$.
Here $K$ is related to the anisotropic parameter $\Delta$,
$K=[1-\cos^{-1}(\Delta)/\pi]^{-1}$. $K=1$ for the isotropic case.
This residual entropy perturbed by the bulk correlation
gives rise to the singular temperature dependence discussed above. 
In this sense, although the structure of the BA equation (\ref{baeh})
and the resultant temperature dependence of eq.(\ref{chi2})
are analogous to the underscreening Kondo effect, 
the underlying physics is a bit different.

\section{Summary}

In this paper, we have given a brief overview of universal aspects of 
the Kondo problem on the basis of the BA exact solution and
boundary conformal field theory.
The finite-size spectra
for various solvable models related to the Kondo problem
obtained from the BA method are consistent with
the results of the CFT analysis.
The exact FSS and the finite-size scaling argument provide 
the exponents of boundary scaling fields characterizing
critical properties of the Kondo problems.
We have also discussed an intriguing relation between the Kondo problems and 
some impurity problems in one-dimensional quantum systems.
It has been shown that a non-magnetic impurity in quantum spin chains
gives rise to a phenomenon similar to the underscreening Kondo effect.

\acknowledgement
A part of this work was carried out under the collaboration with 
Sebastian Eggert, and the late Sung-Kil Yang.
This work was partly supported by a Grant-in-Aid from the Ministry
of Education, Science, Sports and Culture, Japan.

\end{document}